\documentclass[lettersize,journal]{IEEEtran}
\usepackage{amsmath,amsfonts}

\usepackage{algorithm}
\usepackage{array}
\usepackage[caption=false,font=normalsize,labelfont=sf,textfont=sf]{subfig}
\usepackage{textcomp}
\usepackage{stfloats}
\usepackage{url}
\usepackage{verbatim}
\usepackage{graphicx}
\usepackage{cite}
\usepackage{tabularx}
\usepackage{makecell}
\usepackage[top=2cm, bottom=2cm, left=2cm, right=2cm]{geometry}
\usepackage{algorithm}
\usepackage{algorithmicx}
\usepackage{algpseudocode}
\usepackage{amsmath}
\usepackage{makecell}
\usepackage{url}
\usepackage{cite} 
\usepackage[numbers,sort&compress]{natbib}
\usepackage{graphicx}
\usepackage{bbding}
\usepackage{graphicx}
\usepackage{subfig}
\usepackage{overpic}
\usepackage{float} 

\usepackage[misc]{ifsym}
\begin{document}

\title{RA-ICM: A Novel Independent Cascade Model Incorporating User Relationships and Attitudes}


\author{Xinyu Li, Yutong Guo, Jixuan He, Jiacheng Zhao, Chenwei Wang\\

\thanks{Xinyu Li, Yutong Guo, Jixuan He, Jiacheng Zhao, and Chenwei Wang, School of Computer and Information, Hefei University of Technology, Hefei, 230601, China.}
}

\maketitle

\begin{abstract}
The rapid development of social networks has a wide range of social effects, which facilitates the study of social issues. Accurately forecasting the information propagation process within social networks is crucial for promptly understanding the event direction and effectively addressing social problems in a scientific manner. The relationships between non-adjacent users and the attitudes of users significantly influence the information propagation process within social networks. However, existing research has ignored these two elements, which poses challenges for accurately predicting the information propagation process. This limitation significantly hinders the study of emotional contagion and influence maximization in social networks. To address these issues, by considering the relationships between non-adjacent users and the influence of user attitudes, we propose a new information propagation model based on the independent cascade model. 
Experimental results obtained from six real Weibo datasets validate the effectiveness of the proposed model, which is reflected in increased prediction accuracy and reduced time complexity. Furthermore, the information dissemination trend in social networks predicted by the proposed model closely resembles the actual information propagation process, which demonstrates the superiority of the proposed model.

\end{abstract}


\section{Introduction}
\IEEEPARstart{T}{he} continuous development of online social networking services has significantly facilitated people's daily lives. The channels for information sharing and communication among people are consistently expanding. People can now express their views and stances on specific topics or events online, anytime and anywhere. Consequently, the delay in information dissemination caused by geographical and spatial constraints has been greatly reduced. Online social networking platforms, such as Sina Weibo, Twitter, and Facebook, have garnered a massive user base by capitalizing on their features of mobile socialization and real-time information updates. However, the rapid dissemination of various types of information inevitably leads to polarization of people's stances, accompanied by the emergence of a substantial amount of negative information, including rumors and fake news. These phenomena result in diverse social effects that constantly surface, such as: threatening the legitimacy and credibility of online platforms, disrupting social order, influencing the fairness of elections\cite{allcott2017social}, impacting the stock market\cite{difonzo1997rumor}, and even jeopardizing national stability. Consequently, accurately describing the trends in information dissemination on social networking platforms has become an ongoing research focus, which is crucial for the timely understanding of the direction of events and effectively addressing various social effects.

As an integral facet of our daily lives, online social networks have attracted increasing attention and research from scholars. The current research on social networks primarily focuses on information diffusion models and studies on influence maximization. These studies initially relied upon two fundamental models, namely the Independent Cascade (IC) Model\cite{kempe2005influential} and the Linear Threshold (LT) Model\cite{kempe2003maximizing}, to predict the process of information propagation. Nonetheless, with the continuous evolution of social networking platforms, people have discovered that besides neighboring nodes, various factors such as time, topic, space, user emotions, and individual preferences also influence the process of information propagation. Consequently, an extensive body of research has emerged, building upon the foundational IC Model, while taking into account the influence of time \cite{guille2012predictive, haldar2023temporal, chen2012time, kim2014ct}, topic \cite{barbieri2013topic, qin2021influence, 9484767, tian2020deep, michelle2016topic, zhang2020nsti}, spatial dependencies \cite{chen2017modeling}, user emotions \cite{wang2016emotion, wang2017emotion}, and individual preferences \cite{zhang2020nsti, 10339891, wang2020topic, 10443215, dai2022opinion}. This approach aims to make the propagation process of the model more closely resemble the actual trends in information dissemination in social networks.

\begin{figure}[t]
	\centering
	\subfloat[Traditional network model.]{\includegraphics[width=0.33\textwidth]{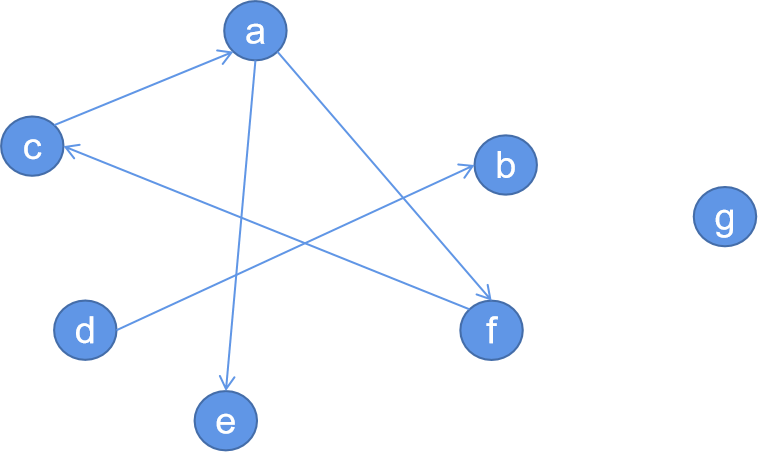}}\\
	\subfloat[Network model of this Paper.]{\includegraphics[width=0.33\textwidth]{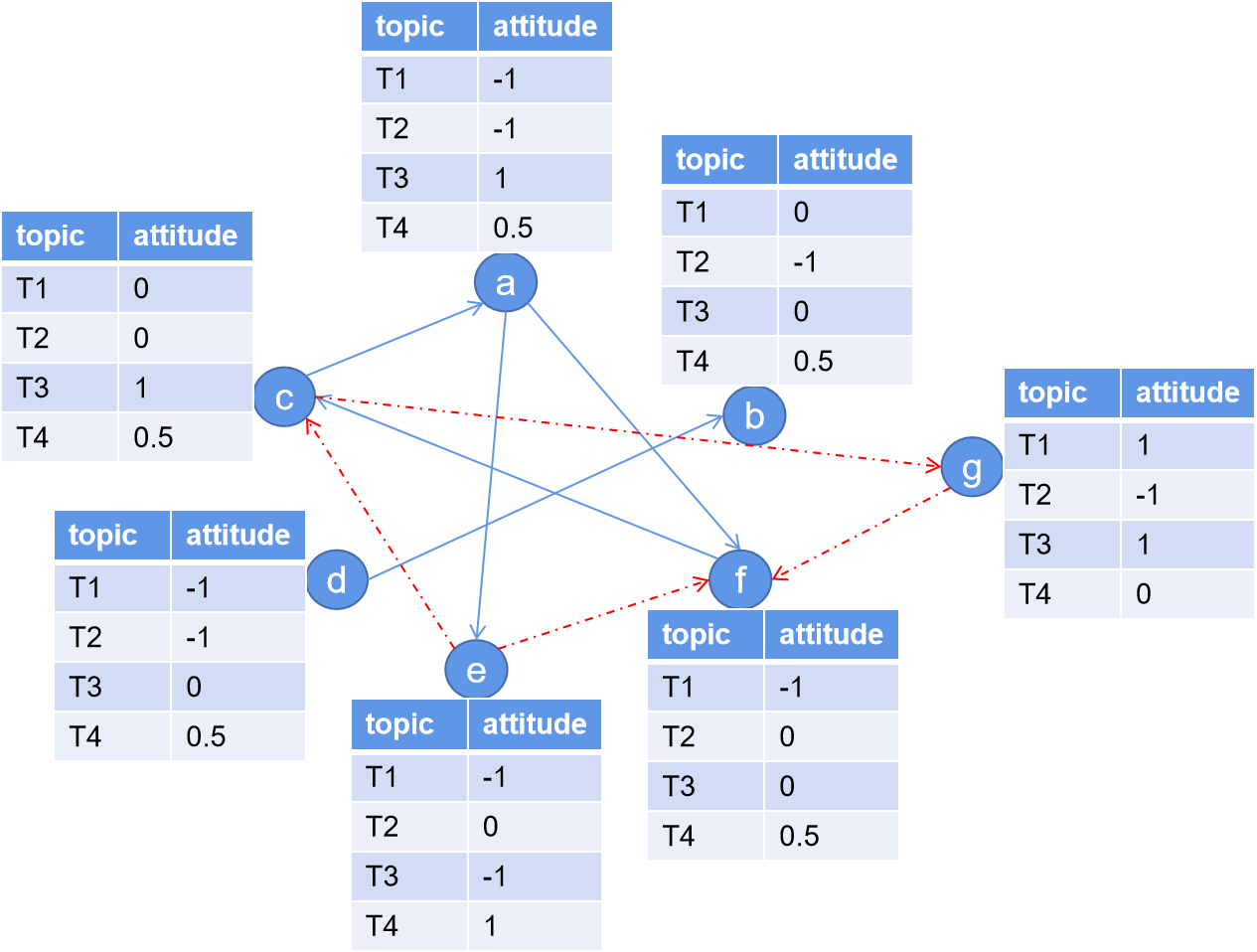}}
	\caption{Network Model.}
    \label{fig:1}
\end{figure}

However, previous studies only considered the influence of neighboring users on the propagation process. It is evident that the information posted by users on social networking platforms like Weibo may not only reach their neighboring users but also a significant number of non-neighboring users. Taking the Weibo platform as an illustration, neighboring users refer to those who have a following relationship with a particular user, while non-neighboring users are those who do not have a following relationship with that user. For example, when a user uses a certain product and shares his experience on a social platform, not only the user's neighboring users can receive that information, but also numerous non-neighboring users who possess a higher interest similarity with the user or exhibit an interest in the product. Furthermore, on social platforms, the information posted by users implicitly reflects their stance on a particular event or product. During the process of information dissemination, the stances of all users towards a specific event or product may undergo constant changes as they are exposed to information with different stances. For example, when a user expresses a supportive stance towards a particular event or product, as the information spreads, some users who initially held opposing or neutral stances may be influenced to shift towards a neutral or supportive stance, while the user's stance may also change to neutral or even opposing, contingent on the various information he receives. \cite{huang2018information} suggests that users' stances also exert a certain influence on the rate of information propagation. However, prior studies have not taken into account the influence of non-neighboring users and the dynamic changes in user stances on the process of information propagation. This omission has had a certain impact on the accuracy of model predictions and, to some extent, has limited research on influence maximization and emotional contagion.

To address the aforementioned issues, this paper undertakes the following work: Firstly, unlike previous approaches, this study models the social network as a directed graph that incorporates topics and user stances, as depicted in Fig.~\ref{fig:1}. Each user is associated with the topics they follow and their stances on these topics. Additionally, it considers the possibility of information propagation between non-neighboring users, such as between user c and user g, as well as between user e and user a. Secondly, this paper improves the mechanism for evaluating the influence between users. In addition to considering the relationship between users, it also takes into account the similarity of the topics they follow and the influence of their stances on these topics. As shown in Fig.~\ref{fig:1}, both user c and user a express interest in topics T3 and T4, and they hold exactly the same stance on these topics. On the other hand, user f and user a are also interested in topics T3 and T4, but they have completely opposing stances on topic T3. Therefore, the influence between user c and user a surpasses that between user f and user a. However, previous influence assessment mechanisms have overlooked the influence of both topic similarity and user stances, which has compromised the accuracy of influence assessment results between users. Thirdly, this paper extends the independent cascade model by considering information propagation between non-adjacent users. As shown in Fig.~\ref{fig:1}, even though there is no direct connection between user C and user G, they both share common interests in topics T1, T3, and T4. Additionally, they hold the same stance regarding topics T2 and T4. Therefore, there is a possibility of information transmission between them. In previous network models, it was assumed that there is no information transmission between user c and user g, which clearly contradicts the reality. Finally, we summarize the contributions of this paper as follows:

\begin{enumerate}
\item{We propose a novel mechanism for calculating user influence that enhances the accuracy of user influence assessment by integrating topic similarity and user stance.}
\item{We extend the independent cascade model to include information propagation between non-adjacent users, providing a more realistic depiction of the information dissemination process in social networks.}
\item{We propose a computational mechanism for measuring user stance changes based on the different stances encountered by users during the information propagation process regarding the same topic. This mechanism dynamically represents the continuous evolution of user stances throughout the information propagation process, effectively improving the model's accuracy in predicting the information dissemination process.}
\item{We constructed six real social network datasets and conducted experiments on these datasets. The experimental results confirmed the authenticity and effectiveness of the proposed model in this paper.}
\end{enumerate}

The remaining sections of this paper are organized as follows: Section \uppercase\expandafter{\romannumeral2} presents the related work. Section \uppercase\expandafter{\romannumeral3} formalizes the computation of user influence, the mechanism of user stance changes, and the process of information propagation and diffusion. Building upon these formalizations, Section \uppercase\expandafter{\romannumeral4} proposes an algorithm for topic and stance awareness model of information propagation and diffusion in social networks. In Section \uppercase\expandafter{\romannumeral5}, a comprehensive analysis is conducted on the experiments performed and their corresponding results. Finally, Section \uppercase\expandafter{\romannumeral6} concludes this paper.

\section{Related Work}
Understanding the propagation mechanisms behind vast amounts of information is crucial in various research fields, including viral marketing\cite{kempe2003maximizing, leskovec2007dynamics}, social recommendations\cite{backstrom2011supervised, xu2014learning}, community detection\cite{fortunato2010community, fortunato2007resolution}, and social behavior prediction\cite{ma2017social, xu2016taxi, zhao2016exploring}. Researchers from diverse fields, such as epidemiology, computer science, and sociology, have conducted extensive studies and proposed various models of information diffusion to accurately describe and simulate this intricate process.

Kempe et al. originally introduced two fundamental models of information diffusion: the Independent Cascade Model (IC)\cite{kempe2005influential} and the Linear Threshold Model (LT)\cite{kempe2003maximizing}. These models provide explanations of the information propagation process in social networks from the perspectives of probability and thresholds. They are widely recognized as the most classic and extensively employed models for describing information propagation in social networks. The IC model assumes that a node u activates its neighboring node v with a probability of p(u,v). The activation behavior between nodes is independent. In this model, any active node in the network has a single opportunity to activate its neighbors, regardless of the outcome. After the propagation process, the node no longer maintains any influence. In the LT model, the number of newly activated nodes in each propagation process depends on the current number of active nodes. This model shares similarities with the IC model in terms of the propagation process. However, it differs in one aspect. Unlike the IC model, which assumes that the activation behavior between nodes is mutually independent, this model assumes that in each propagation process, every unactivated node can determine its activation state by comparing the sum of the linearly weighted values of all its activated neighboring nodes with the corresponding threshold. Furthermore, in the IC model, each node has only one chance to activate its neighboring nodes, whereas, in the LT model, each activated node has multiple opportunities to activate other unactivated nodes.

With the rapid development of social networks, researchers have discovered that in addition to neighboring nodes, various factors such as time, topic, space, user emotions, and individual preferences also exert influence on the process of information propagation. Consequently, traditional information propagation models are no longer capable of accurately describing the process of information dissemination in social networks. As a result, extensive research has been conducted to incorporate the aforementioned factors into the IC Model, resulting in certain improvements to traditional information propagation models. These modifications are aimed at enhancing the predictive accuracy of the models, thereby enabling them to closely resemble the actual process of information dissemination.

Taking into account the influence of time, \cite{guille2012predictive} proposed T-BaSIC, an asynchronous information propagation model based on the IC Model. T-BaSIC considers temporal effects in the information dissemination process and provides an explanation for the dynamic propagation of information at the macro level by considering micro-level interactions and interconnected topology among users. T-BaSIC has demonstrated effective prediction of the dynamic process of information propagation. \cite{haldar2023temporal} introduces the Temporal Independent Cascade model (T-IC), an extension of the IC Model that incorporates temporal characteristics. This model effectively captures the temporal aspects of the network, such as the duration of evolving connections or the dynamic likelihood of propagation along connections, thereby improving the predictive performance of the model. \cite{chen2012time} extends the IC Model by incorporating the time factor of information diffusion in social networks. They propose the Independent Cascade Model with Meeting Events (IC-M), a new propagation model that captures the temporal delays in information propagation. \cite{kim2014ct} extends the IC Model by considering that in real propagation processes, users have multiple opportunities to activate neighboring nodes. They propose a new and practically meaningful Continuous Activation and Time-Limited Influence Diffusion Model (CT-IC). This model assumes that each active node repeatedly activates its neighboring nodes until a given time, thereby improving the predictive accuracy of the model. \cite{li2013modeling} introduces the GT propagation model by considering the influence of time on user behavior in the process of information propagation. This model incorporates both time-related and non-time-related information into the decision-making process of users, enhancing the accuracy of modeling and predicting the process of information diffusion. It effectively forecasts changes in user behavior decisions within specific time intervals.

Taking into account the influence of topics, researchers in \cite{barbieri2013topic, 9613773} investigate the impact of user topic dependency on the information propagation process, building upon the IC model. They enhance the evaluation method for assessing the infection probability between users and propose a novel topic-aware influence-driven propagation model, which provides a more precise depiction of cascade effects in real social networks. \cite{qin2021influence} introduces a topic-aware community-based independent cascade model \cite{10251628} that incorporates both the characteristics of community structure and the integration of community topic features. This model demonstrates enhanced stability, dynamic adaptability, increased computational efficiency, and reduced space requirements, all while preserving the approximation ratio and influence scope. Furthermore, \cite{tian2020deep} presents a topic-aware social influence propagation model inspired by viral marketing, based on the IC model. This model integrates topic awareness, facilitating more personalized and precise advertising. In \cite{michelle2016topic}, drawing inspiration from the PageRank algorithm, the researchers propose a topic-sensitive information diffusion model known as TS-IDM (topic-sensitive information diffusion model). The model measures and quantifies the interests of individual users using relevance scores. These scores, in conjunction with an evolutionary game model of information diffusion, are employed to enhance the accuracy of information propagation across various topics. \cite{zhang2020nsti} by combining the influence of neighboring structures on different topics with the distribution of user interests across various topics, a propagation model called NSTI-IC (neighborhood structure and topic-aware interest) is proposed. This model improves the predictive accuracy of information propagation.

Taking into account the influence of spatial factors, \cite{qin2021influence} introduces a community-based independent cascade model based on the IC model, considering the issue of spatial consumption in the information propagation process. This model incorporates the spatial structural features of communities into the IC model, reducing the complexity of dynamic information propagation without sacrificing accuracy. Considering specific spatial dependencies, \cite{chen2017modeling} introduces a spatial Markov-dependent influence propagation model based on the existing IC model. This model offers an improved depiction of the information diffusion process in online social networks.

Taking into account the influence of user emotions, \cite{wang2016emotion} introduces an emotion-based independent cascade model, recognizing that emotions can contribute to the broader dissemination of information. This model illustrates the process of information propagation in online social media influenced by emotional contagion. In this model, information is represented by the distribution values of five different emotions. These emotions have a certain degree of influence on the magnitude of influence between users. The propagation probability of each piece of information is computed as the weighted average of the diverse emotions. This model enhances predictive performance and offers insights for applying emotional contagion in viral marketing on online social media platforms.Considering the dissemination of emotions in the process of information propagation, \cite{wang2017emotion} introduces a sentiment-based independent cascade model to examine the dynamics of emotional contagion. A learning model is developed that incorporates user features, structural features, and tweet features to predict changes in emotions after retweeting. The model calculates conversion weights to anticipate the emotional transformation outcomes of retweets.  This model exhibits superior performance compared to conventional information propagation models.

Taking into account the influence of individual preference factors, \cite{zhang2020nsti} integrates the impact of individual preferences into the propagation process of the model, proposing a propagation model called NSTI-IC. This model enhances prediction accuracy by combining structural influence and topic-aware interests. \cite{wang2020topic} has developed a unified probabilistic framework that formalizes the problem as a topic-enhanced sentiment propagation model. This model predicts users' sentiment states based on their historical emotional states, the topic distribution of tweets, and the social structure. It provides a more precise description of users' emotional prediction changes during the information propagation process. \cite{dai2022opinion} proposes a dynamic enhanced independent cascade (EIC) model based on the IC model, taking into account both group polarization effects and individual preferences. The model utilizes real data to detect and model the information propagation mechanism influenced by group polarization effects. It effectively captures the impact of group polarization effects and individual preferences on the information propagation process.

In addition, there are some related studies as follows. \cite{xia2019novel} proposes a particle dynamics-based information propagation model by considering the initial influence of information at the start of propagation and the activity of users when retweeting information. \cite{mashayekhi2018weighted} investigates the learning problem of diffusion probabilities in the IC model and presents a weighted estimation method for diffusion probabilities, which reduces the average prediction error of the model. \cite{chen2011influence} introduces the IC-n model with negative opinions based on the IC model, incorporating the diffusion of both positive and negative information. This model captures the evolution and propagation process of negative information during the information dissemination process. \cite{yang2023triadic} proposes the TC-IC model based on the IC model, which utilizes the triadic closure structure of the social network to accurately measure the closeness between nodes. By assigning different propagation probabilities to each edge, this model significantly enhances the efficiency of identifying influential nodes. \cite{qiu2022best} proposes the BHICM algorithm, an improved heuristic approach. It introduces a dynamic correlation strategy between propagation probabilities and hop counts, eliminating the need to process neighboring nodes of seed nodes. There are also other propagation models, such as the Susceptible-Infected-Recovered (SIR) model\cite{hethcote2000mathematics}, the Susceptible-Infected-Susceptible (SIS) model\cite{zafarani2014social}, the Twin-SIR spreading model\cite{yi2021research}, the Susceptible-Infected-Recovered-Anti-spreader model\cite{kumar2023sira}, and the SIR-IM model\cite{qiu2021sir}. Most of these models are infectious models and assume that information spreads from a set of source (or seed) nodes. Other nodes can only access information through their neighboring nodes. During the information propagation process, nodes in the network have only two states: either active or inactive. Moreover, active nodes can influence their neighboring nodes through the execution process of the propagation model. Over time, the number of active nodes will gradually increase\cite{barbieri2013topic}. However, as pointed out in\cite{prakash2012threshold}, the IC diffusion process is more accurate in describing the information propagation process in social networks compared to infectious disease models like SIR. Therefore, this paper primarily focuses on the information diffusion model in social networks from the perspective of the IC model.

However, the aforementioned studies have overlooked the relationships between non-adjacent users and the role of user stances in the information propagation process of social networks. This oversight significantly impacts the predictive accuracy of the models. To address this gap, this paper proposes a new information propagation model that takes into account the influence of these two factors. The model not only improves prediction accuracy but also reduces time complexity. Consequently, the model's predictions align more closely with the actual information propagation process observed in real social networks.

\section{PROBLEM FORMULATION}
For contemporary online social networking platforms, the information that users can receive is no longer limited to the scenarios described by previous models: where nodes are influenced solely by their previously active neighboring nodes or solely influenced by their neighboring nodes. However, in reality, a node's stance on a particular topic can be influenced not only by its neighboring nodes but also by non-adjacent nodes that share similar topic interests, albeit with a certain probability. Furthermore, each node has multiple opportunities to influence other nodes, rather than just once. Within a certain time frame, nodes possess the ability to impact other nodes. In this section, the paper first introduces the network model used in this study. Next, it presents the measurement methods for assessing influence between nodes, including topic similarity measurement among users, calculation of the influence between users, determination of the stance persistence of nodes towards a certain topic, and the approach to determining the node's state in a specific topic. Finally, we provide a comprehensive description of the topic and stance-aware social network diffusion model.

\subsection{Network Model}
As shown in Fig.~\ref{fig:1}, in the proposed topic and stance aware information diffusion model, the social network is modeled as a directed graph $G = (\boldsymbol{V},\boldsymbol{E},\boldsymbol{T})$, where $\boldsymbol{V}$ is the set of nodes of size n, $\boldsymbol{E}$ is the set of edges of size m, and $\boldsymbol{T}$ is the set of topics in the network. Here, nodes represent users in online social networks, edges represent connection relationships between users, and edge $e_{ij}$ represents the direction of information flow from node i to j. Each node $v_{i}\in \boldsymbol{V}$ corresponds to a set of topics $\boldsymbol{T_{i}}=\{t_{i\_1},t_{i\_2},t_{i\_3},……,t_{i\_z}\}$. The set reflects the attitude of user i towards the information of all topics, where  $t_{i\_k} \in \{-1,0,0.5, 1\}$,$t_{i\_k}$  is used to represent user i's attitude towards topic k. The value -1 indicates that user i has not yet encountered topic k, meaning the user's stance on topic k is unknown. The value 0 represents user i's opposition to topic k, while the value 0.5 represents a neutral stance. A value of 1 indicates user i's agreement with topic k. These three states represent the known stance of the user regarding topic k. The variable z represents the total number of topics in the social network. Tabel~\ref{tab1} summarizes the commonly used symbols and their meanings.

\begin{table*}[t]
\caption{FREQUENTLY USED NOTATION\label{tab1}}
\centering
\begin{tabularx}{\textwidth}{ c
         >{\raggedright\arraybackslash}X
        }
\hline
Notation & Description \\
\hline
$G=(\boldsymbol{V},\boldsymbol{E},\boldsymbol{T})$ & Represents a social network, where $\boldsymbol{V}$ is a node set of size n, $\boldsymbol{E}$ is an edge set of size m, where edge $e_{ij}$ represents the direction of information flow from node i to j, and $\boldsymbol{T}$ is the set of topics in the network.\\

$P_e$ & Probability of information propagation on edge e\\

$n = |\boldsymbol{V}|$ & The number of nodes in G\\

$m =|\boldsymbol{E}|$ & The number of edges in G\\

$\boldsymbol{T_i}=<t_{i\_1},t_{i\_2},……,t_{i\_z}>$ & Attitude held by user i towards information on all topics\\

$z = |\boldsymbol{T}|$ & The number of topics in G\\

$P_{uv}{i}$ & Influence of u on its neighbor node v against topic i\\

$P_{uv}{~i}$ & Influence of u on a node v among its adjacent nodes that is unknown to the topic ti\\

$P_{~uv}{i}$ & Influence of u on its non-adjacent node v against topic i\\

$P_{~uv}{~i}$ & Influence of u on node v, which is unknown to topic ti among its non-adjacent nodes\\

$sim(u,v)$ & Interest similarity between nodes u and v\\

$\delta, \lambda, \mu, \epsilon$ & Control parameter\\

$A_{v}{i}$ & Attitude value of node v for topic i\\
\hline
\end{tabularx}
\end{table*}

\subsection{Influence Evaluation between Nodes}
In online social networks, there are two main sources of information that influence the views and stances of the target user v. These sources come from both the neighboring nodes of the target user and non-neighboring nodes. Furthermore, for both neighboring and non-neighboring nodes, there are two scenarios of influence: when the node u shares the same topic with the target user v, and when the node u possesses a topic that is unknown to the target user v.

Definition 1: Topic similarity between users

Given two nodes u, v, the topic sets corresponding to these two nodes are $\boldsymbol{T_{u}}=\{t_{u\_1},t_{u\_2},t_{u\_3},……, t_{u\_z}\}$ and $\boldsymbol{T_{v}}=\{t_{v\_1},t_{v\_2},t_{v\_3},……,t_{v\_z}\}$, then the topic similarity sim(u,v) between node u and node v is measured as follows.

\begin{equation}
sim(u,v) = 1/\bigg( \Big( \sqrt{z}+\sqrt{\sum_{i=1}^z\big( t_{u}^{i} - t_{v}^{i}\big)^2}\Big)/\sqrt{z}\bigg) \label{eq:1}
\end{equation}

Where $sim(u,v)\in[1/\sqrt{2},1]$, a larger sim(u,v) indicates a more significant topic similarity between node u and node v.

Definition 2: The size of influence between users

Given two nodes u and v, the influence $P_i$ of node u on node v on topic $t_i$ is measured as follows:

\begin{equation}
P_{i}(u,v) = \delta \ast sim(u,v) \ast f(t_{v}^{i},t_{u}^{i}) 
\label{eq:2}
\end{equation}

\begin{equation}
f(t_{v}^{i},t_{u}^{i})=\left\{
	\begin{array}{rcl}
	1 ,& & t_{v}^{i} = -1,0.5 or t_{v}^{i} = t_{u}^{i}\\
	\lambda, & & |t_{v}^{i} - t_{u}^{i}| \le 0.5\\
	\mu ,& & else\\
	\end{array}
	\right.   
\label{eq:3}
\end{equation}

In the Eq.(\ref{eq:2}), $\delta$ is a parameter with a value range of [0,1], which is used to control the influence weight of user u to user v. When there is edge $e_{uv}\in \boldsymbol{E}$, the value range of a should be [0.5,1], and when $e_{uv}\notin \boldsymbol{E}$, the value range of $\delta$ is set to [0,0,5). Due to the influence of current state-of-the-art social network recommendation algorithms, there is a certain probability that user v will come across information from user u, regardless of whether there is a direct relationship between user u and user v. Specifically, when there is a relationship between them, the likelihood of user v encountering information from user u is higher. However, it does not imply that user v will always receive information from user u just because they have a relationship. Therefore, in this case, the range of $\delta$ is set to (0.5,1]. When there is no relationship between user u and user v, there is still a possibility for user v to encounter information from user u. However, compared to users who have a relationship, the probability of encountering the information is relatively smaller. Therefore, the value range of $\delta$ is set to [0,0.5). sim(u,v) represents the topic similarity between users u and v. In the Eq.(\ref{eq:3}), $\lambda$ is a parameter with a value range of [0.5,1], and $\mu$ is a parameter with a value range of [0,0,5). These two parameters are used to control the weight of the influence of user u on user v under topic i. In other words, regardless of whether there is a connection between two nodes or the topic similarity, there is a certain probability of mutual influence, which is different from previous models that only consider information propagation among adjacent nodes. Clearly, it can be observed that when the similarity between non-adjacent nodes is significantly greater than the similarity between adjacent nodes, the influence of non-adjacent nodes on the target node is not necessarily smaller than the influence of adjacent nodes on the target node.

In particular, in order to facilitate the subsequent algorithm analysis, we define that when u is the adjacent node of v, if user u and target user v have the same topic $t_i$, then the size $P_{i} (u,v)$ that user v will be influenced by user u under the topic $t_i$ is expressed as $P_{uv}^{i}$. If user u has the unknown topic $t_i$ of user v, The size $P_{i} (u,v)$ that user v will be influenced by user u under topic $t_i$ is denoted as $P_{uv}^{~i}$. When u and v are non-adjacent nodes, if user u and target user v have the same topic $t_i$, the size $P_{i} (u,v)$ that user v will be influenced by user u under topic $t_i$ is expressed as $P_{~uv}^{i}$. If user u has unknown topic $t_i$ of user v, the size $P_{i} (u,v)$ will be influenced by user u. The size $P_{i} (u,v)$ that user v will be influenced by user u under topic $t_i$ is denoted as $P_{~uv}^{~i}$.

Definition 3: Node's stance persistence to a topic

For any topic $t_i$ in the social network G, node v has a stance persistence $A_{v}^{i}$ towards that topic. This value is used to indicate the likelihood of user v's stance changing on that topic, as follows:

\begin{equation}
A_{v}^{i} = A_{v}^{i} - \sum_{u=1}^k\Big(|t_{u}^{i} - t_{v}^{i}| \ast P_{i}(u,v) - (\overline{t_{u}^{i} \oplus t_{v}^{i}}) \ast P_{i}(u,v) \Big)/k
\label{eq:4}
\end{equation}

In the Eq.(\ref{eq:4}), k represents the number of information received by node v so far, $A_{v}^{i}$ represents the stance persistence of node v to topic $t_i$, with a range of values between [0, 1]. A higher value of $A_{v}^{i}$ indicates that node v's current stance on topic $t_i$ is less likely to be influenced by surrounding users.  It is evident that node v's stance persistence $A_{v}^{i}$ towards the current topic $t_i$ is influenced by receiving information with different stances on that topic. When node v receives information with different stances on topic $t_i$, $A_{v}^{i}$ decreases, indicating an increased likelihood of node v's stance changing on topic $t_i$. Conversely, when node v receives information with the same stance on topic $t_i$, $A_{v}^{i}$ increases, reducing the likelihood of node v's stance changing on topic $t_i$.

Definition 4: Node status in a topic.

For a given node v, when it receives information from other nodes regarding a certain topic $t_i$, the stance of node v on that topic may change. The measurement criteria for assessing this change are as follows:

When node v is unknown or neutral about a topic i, the state of node v at the next moment is as follows.

\begin{equation}
t_{v}^{i}=\left\{
	\begin{array}{rcl}
	t_{u}^{i} ,& & P_{i}(u,v)\ge A_{v}^{i}\\
	0.5, & & else\\
	\end{array}
	\right.   
\label{eq:5}
\end{equation}

When node v has an unknown or neutral stance towards topic i, receiving information on that topic can lead to two possible changes in its stance. If the influence of node u on node v regarding topic i is greater than the stance value $A_{v}^{i}$ of node v, then node v's stance of topic i becomes equal to node u's stance of topic i. However, if the influence is smaller than $A_{v}^{i}$, node v's stance on topic i remains neutral, that is, $t_{v}^{i}$ equals 0.5.

When node v is in the support or opposition state for topic i, the state of node v at the next moment is as follows.

\begin{equation}
t_{v}^{i} = (\overline{t_{u}^{i} \oplus t_{v}^{i}}) \ast  t_{v}^{i} + (t_{u}^{i} \oplus t_{v}^{i}) \ast (t_{v}^{i} \pm \varepsilon \ast 0.5)
\label{eq:6}
\end{equation}

As shown in the Eq.(\ref{eq:6}), where $\varepsilon$ is a constant parameter, when $P_{i}(u,v) \textgreater A_{v}^{i}$, $\varepsilon$ is 1, when $P_{i}(u,v) \textless A_{v}^{i}$,$\varepsilon$ is 0,$\pm$ is used to control the direction of stance change of node v. When the initial $t_{v}^{i}$ is 0, indicating that node v initially holds an opposing stance on topic $t_i$, we assign a positive sign (+). Conversely, when the initial $t_i$ is 1, indicating that node v initially holds a supportive stance on topic $t_i$, we assign a negative sign (-).

\subsection{Diffusion Models in Social Networks based on Topic and Stance Awareness}
In a social network, each node can be in one of four states - unknown, opposing, neutral, or supporting - for any of the n topics at any given time. Initially, users who hold opposing, neutral, or supporting stances on a particular topic are considered active users for that topic, while nodes with unknown stances are considered inactive users for that topic. These active users have the opportunity to influence other nodes in the social network with a certain probability, regardless of whether they are connected or not. The state of other nodes in relation to a specific topic can be influenced by the active user's stance on that topic. Although the model in this paper is based on topic and stance awareness, our approach differs from existing topic-aware works\cite{barbieri2013topic, qin2021influence, tian2020deep, michelle2016topic, zhang2020nsti}. They assume that the activation probability on each edge (u, v) is determined by the similarity of topic distributions between nodes u and v. Furthermore, they assume that each node can only be influenced by its neighboring nodes, and the neighboring nodes are guaranteed to have an influence on the node. However, in real social networks, the probability of influence between nodes u and v depends not only on the similarity of their topic distributions but also on the differences in their stances on specific topics. Clearly, the probability of influence between nodes with opposing opinions is lower than the probability of influence between nodes with similar opinions. Additionally, for neighboring nodes, node v is not guaranteed to receive information from those neighboring nodes. Moreover, whether node v is activated depends not only on its neighboring nodes but also on the probability of being influenced by other non-neighboring nodes, regardless of whether they share the same topic with node v.

In each step, we assume that all nodes have the ability to activate their neighboring nodes or non-neighboring nodes at any given time. We make this assumption for the following reasons: Firstly, we assume that the information users encounter in a social network is not limited by time constraints. In the current social networking environment, social networks recommend information to users from any period. Therefore, continuing to use the previous models where all nodes have the ability to influence their neighboring nodes only in the next time step after being activated does not accurately reflect the information propagation process in the current social networking environment. Secondly, we assume that all nodes not only receive information from their neighboring nodes but also have a certain probability of receiving information from their non-neighboring nodes. Because in current social networks such as Weibo, Facebook, and others, users can not only receive information from their connected users but also come across information from completely unrelated users. The propagation models proposed in previous models, where nodes can only influence their neighboring nodes and can only be influenced by their neighboring nodes, are not well-suited for the current online social networks \cite{10149418}.

We take node a of the network model in Fig.~\ref{fig:1} as an example, specifically describe the propagation process of the proposed social network diffusion model based on topic and stance awareness. At time 0, the topic set corresponding to node a is $\boldsymbol{T_{a}}=\{-1, 0, 1, 0.5\}$. At this time, node a will be affected by node c. For unknown topic t1, the influence value of node c on node a is $P_{ca}^{~1} = P_{1}(c,a)=\delta \ast \bigg(1/\Big(\big(\sqrt{\sum_{i=1}^4(t_{a}^{1}-t_{c}^{1})^2} + 1e^{-10}\big)/\sqrt{4}\Big)\bigg)$,for a known topic $t_3$, the influence value generated by node c on node a is $P_{ca}^{3} = P_{3}(c,a)=\delta \ast \bigg(1/\Big(\big(\sqrt{\sum_{i=1}^4(t_{a}^{3}-t_{c}^{3})^2} + 1e^{-10}\big)/\sqrt{4}\Big)\bigg) \ast \lambda$. Influenced by node c, the stance adherence of node a to topic $t_1$ is $A_{a}^{1} = A_{a}^{1} - (|t_{a}^{1} - t_{cc}^{1}| \ast P_{1}(c,a))$,the stance adherence of node a to topic $t_3$ is $A_{a}^{3} = A_{a}^{3} - (0 - P_{3}(c,a))$.In this case, for topic $t_1$, if $P_{1}(c, a)$ is greater than $A_{a}^{1}$, then $t_{a}^{1}$=$t_{c}^{1}$; otherwise, $t_{a}^{1}$=0.5, indicating that node a is already in a known state for topic $t_1$, but it is not affected by node c's stance toward topic $t_1$ and remains neutral.For topic $t_3$,$t_{a}^{3} = (\overline{t_{a}^{3} \oplus t_{c}^{3}}) \ast t_{a}^{3} = t_{a}^{3}$,indicating that the stance of node a towards topic $t_3$ does not change after being influenced by node c. Meanwhile, node a can also be influenced by other non-adjacent nodes with probability p, such as nodes b and g, and the influence process is the same as above.

\section{Algorithm}
In this section, we will provide a detailed description of the proposed topic and stance-aware social network diffusion model algorithm. Unlike previous algorithms, the model proposed in this paper is insensitive to time and considers the influence of non-neighboring nodes. Additionally, it incorporates the calculation of user stances on topics, taking into account the impact of user stances on the diffusion process. The detailed procedure of the algorithm is as follows.

\subsection{Information Dissemination Process}
Different from traditional information diffusion models, we propose a Topic and Stance Aware social network diffusion model based on the current landscape of new-generation social networks. Considering that information in the current landscape of new-generation social networks is no longer time-sensitive, meaning that users are exposed to not only the most recent information but also information from previous time points, our model assumes that the information posted by user v at any time $t_{1}$ may be received by user u at a later time $t_{2}$ (where $t_{2}(t_{2} \textgreater t_{1})$). In addition, for users in a social network, the information they are exposed to is not limited to their connected nodes. Therefore, our model assumes that user v can access information not only from its neighboring nodes but also from non-neighboring nodes. Regardless of the topic similarity sim(u,v) between users, user v has a certain probability of accessing information from other users u. The specific propagation process is illustrated in Algorithm $1$.

\floatname{algorithm}{Algorithm}
\renewcommand{\algorithmicrequire}{\textbf{Input:}}
\renewcommand{\algorithmicensure}{\textbf{Output:}}

\begin{algorithm}
    \caption{TSA}
        \begin{algorithmic}[1] 
            \Require $G=(\boldsymbol{V},\boldsymbol{E}, \boldsymbol{T})$, initial set of topics $\boldsymbol{T_{k}}=\{\boldsymbol{T_{1}},\boldsymbol{T_{2}},……,\boldsymbol{T_{z}}\}$,initial topic number j,number of propagation K.
            \Ensure  The updated network G=(V,E,T).
            \Function {TSA}{$G,\boldsymbol{T_{k}}, j, K$}
            \State $\boldsymbol{T_{j-0}^{now}} = \boldsymbol{T_{j-0}}$, $\boldsymbol{T_{j-0.5}^{now}} = \boldsymbol{T_{j-0.5}}$, $\boldsymbol{T_{j-1}^{now}} = \boldsymbol{T_{j-1}}$
            \State $\boldsymbol{V_{j}^{new}} = \boldsymbol{T_{j-0}^{now}} \cup \boldsymbol{T_{j-0.5}^{now}} \cup \boldsymbol{T_{j-1}^{now}}$
            \State $\boldsymbol{V_{adj}} = \boldsymbol{\emptyset}$, $\boldsymbol{S} = \boldsymbol{\emptyset}$
                \For{k = 1 to K}
                    \For{v in $\boldsymbol{V_{j}^{new}}$}
                        \State $\boldsymbol{S}=\{u|(u,v)\in \boldsymbol{E}\}$
                        \For{q $\in$ $\boldsymbol{S}$}
                            \If{q $\notin \boldsymbol{V_{adj}}$}
                                \State $T_{cur} = T_{q}^{j}$
                                \State $A_{q}^{j} = Eq.(4)$
                                \State $T_{q}^{j} = ATT(G,q,v,j,A_{q}^{j})$
                                \State $\boldsymbol{V_{adj}} \gets q$
                            \EndIf
                            \If{$T_{cur} = -1$ and $T_{q}^{j} != -1$}
                                \State $\boldsymbol{T_{j-T_{q}^{j}}} \gets q$
                                \State $\boldsymbol{V_{j}^{new}} \gets q$
                            \EndIf
                        \EndFor
                    \EndFor
                    \State $\boldsymbol{V_{~adj}} \gets \boldsymbol{V}\backslash \boldsymbol{V_{adj}}$
                    \State $NADJ(G,j,\boldsymbol{V_{~adj}},\boldsymbol{V_{j}^{new}})$
                \EndFor
                \State \Return{$G$}
            \EndFunction
        \end{algorithmic}
\end{algorithm}

Here,$\boldsymbol{V}\backslash \boldsymbol{V_{adj}}$ denotes the difference set between the set $\boldsymbol{V}$ and the set $\boldsymbol{V_{adj}}$, for example, $\boldsymbol{V}=\{a,b,c,d,e\}$, $\boldsymbol{V_{adj}}=\{b,d\}$, then $\boldsymbol{V}\backslash \boldsymbol{V_{adj}} =\{a, c,e\}$. In the TSA algorithm, the adjacency node $\boldsymbol{V_{adj}}$ and the non-adjacency node $\boldsymbol{V_{~adj}}$ are comprehensively considered, as well as all the users known and unknown to the specified topic j.

\subsection{Node State Change Process}
During the process of information propagation, the stance of a node towards the current topic may continuously change. We consider the current stance of the node and the influence of all the different stance information it has received up until time t to determine the possible stance changes at the current time t. Clearly, when the node is in an unknown or neutral state, it is susceptible to the influence of different stance information. However, when the node is in a supporting or opposing state for the current topic, the change in stance depends on factors such as the similarity of topics between the node and other users, as well as all the different stance information it has received so far. The specific calculation process is shown in Algorithm $2$.

\floatname{algorithm}{Algorithm}
\renewcommand{\algorithmicrequire}{\textbf{Input:}}
\renewcommand{\algorithmicensure}{\textbf{Output:}}

\begin{algorithm}
    \caption{ATT}
        \begin{algorithmic}[1] 
            \Require $G=(\boldsymbol{V},\boldsymbol{E}, \boldsymbol{T})$, node q,v, the topic number j,$A_{q}^{j}$.
            \Ensure  $T_{q}^{j}$.
            \Function {ATT}{$G,q,v,j,A_{q}^{j}$}
            \State $T_{cur} = T_{q}^{j}$
             \If{$T_{q}^{j}=0.5$ or -1}
                \State $T_{q}^{j} = Eq.(5)$
            \Else
                \State $T_{q}^{j} = Eq.(6)$
            \EndIf
            \If{$T_{cur} != T_{q}^{j}$}
                \State $\boldsymbol{T_{j-T_{q}^{j}}} \gets q$
                \State $\boldsymbol{T_{j-T_{cur}}} \gets q$
            \EndIf
            \State \Return{$T_{q}^{j}$}
            \EndFunction
        \end{algorithmic}
\end{algorithm}

The process of state change for node q regarding topic j can be determined based on Definition 4. It distinguishes the state of node q for a specific topic j. If node q is in an unknown or neutral state for topic j, it can be derived using Eq.~\ref{eq:5}. If node q is in a supporting or opposing state for topic j, it can be derived using Eq.~\ref{eq:6}. Particularly, when the state of node q for topic j changes, the corresponding set also undergoes dynamic changes.

\subsection{Propagation in Non-adjacent Nodes}
Nodes also influence their non-neighboring nodes during the information propagation process. Whether these nodes are in a known or unknown state for the current topic, there is a certain probability that they will be influenced by the information from the node. The specific propagation process is shown in Algorithm $3$.

\floatname{algorithm}{Algorithm}
\renewcommand{\algorithmicrequire}{\textbf{Input:}}
\renewcommand{\algorithmicensure}{\textbf{Output:}}

\begin{algorithm}
    \caption{NADJ}
        \begin{algorithmic}[1] 
            \Require $G=(\boldsymbol{V},\boldsymbol{E}, \boldsymbol{T})$,$\boldsymbol{V_{~adj}},\boldsymbol{V_{j}^{new}}$,the number of propagation k, the topic number j, the parameters r1,r2.
            \Ensure  The updated network $G=(\boldsymbol{V},\boldsymbol{E},\boldsymbol{T})$.
            \Function {NADJ}{$G,\boldsymbol{V_{~adj}},\boldsymbol{V_{j}^{new}},k,j,r1,r2$}
            \State Generating collections $\boldsymbol{V^{new}},\boldsymbol{V_{~adj}^{new}},\boldsymbol{V^{new}}=\{v|v\in \boldsymbol{V_{j}^{new}}\}$
            \State $|\boldsymbol{V^{new}}|=r1\ast|\boldsymbol{V_{j}^{new}}|$
            \State $\boldsymbol{V_{~adj}^{new}}=\{\boldsymbol{U} \cup \boldsymbol{V}|u,v \in \boldsymbol{V_{~adj}},u \in \boldsymbol{T_{j}}, v \notin \boldsymbol{T_{j}}\}$
            \State $|\boldsymbol{V_{~adj}^{new}}|=r2\ast|\boldsymbol{V_{~adj}}|$
            \For{$q \in \boldsymbol{V_{~adj}^{new}}$}
                \State $T_{cur} = T_{q}^{j}$
                \For{$v \in \boldsymbol{V^{new}}$}
                    \State $A_{q}^{j} = Eq.(4)$
                    \State $ATT(G,q,v,j,A_{q}^{j})$
                    \If{$T_{cur} = -1$ and $T_{q}^{j} != -1$}
                        \State $\boldsymbol{V_{j}^{new}} \gets q$
                        \State $\boldsymbol{T_{j}} \gets q$
                    \EndIf
                \EndFor
            \EndFor
            \State \Return{$G$}
            \EndFunction
        \end{algorithmic}
\end{algorithm}

In this context, $\boldsymbol{T_j}$ represents the set of users for whom topic j is known, and $\boldsymbol{V_{~adj}^{new}}$  contains a subset of non-neighboring users for node v. This subset includes both users who have knowledge about topic j and those who are unaware of it. The parameter r2 determines the size of the set $\boldsymbol{V_{~adj}^{new}}$. And for the collection of certain proportion relationship in the $\boldsymbol{U}$ and $\boldsymbol{V}$, the $|\boldsymbol{U}| = r \ast |\boldsymbol{V_{~adj}^{new}}|$, $|\boldsymbol{V}| = a \ast |\boldsymbol{V_{~adj}^{new}}|$, among them, $r\in[0.5,1], a\in[0,0.5),r + a=1$.

\section{Experiment}
In this section, we will perform tests on our algorithm using six real-world datasets obtained through Python web scraping. To showcase the effectiveness and credibility of our propagation method, we will compare it with several existing models. Additionally, we will conduct visual analysis of user activation and stance changes during the information propagation process.

\subsection{Dataset}
We have categorized the dataset into two main categories: datasets containing a single topic and datasets containing multiple topics concurrently. Our study investigates the information propagation process described in our proposed model using six real-world datasets. Datasets containing only a single topic are as follows: Dataset \uppercase\expandafter{\romannumeral1} encompasses all original Weibo posts and Weibo topic comments pertaining to the specific topic of "queue jumping" on Sina Weibo. The data was collected from May 2, 2023, to June 2, 2023. It comprises 1,300 user nodes and 4,951 edges. Dataset \uppercase\expandafter{\romannumeral2} comprises all original Weibo posts and Weibo topic comments related to the topic of accommodation provided by a specific chain of hotels. The data was collected from April 12, 2023, to May 12, 2023. It includes 1,061 user nodes and 4,122 edges. Dataset \uppercase\expandafter{\romannumeral3} contains all original Weibo posts and Weibo topic comments regarding a public figure's involvement in politically sensitive issues. The data was collected from April 12, 2023, to May 12, 2023. It consists of 1,791 user nodes and 5,895 edges.

Datasets containing multiple topics are as follows: Dataset \uppercase\expandafter{\romannumeral4} comprises all original Weibo posts and related comments pertaining to the topics of "queue jumping" and a marketing issue concerning a restaurant during the holidays. The data was collected from April 12, 2023, to June 12, 2023, and includes 2,300 user nodes, 8,780 edges, and 2 topics. Dataset \uppercase\expandafter{\romannumeral5} includes all original Weibo posts and related comments discussing the topics of "budget travel during the holidays" and a public figure's involvement in politically sensitive issues. The data was collected from April 15, 2023, to May 15, 2023, and includes 2,754 user nodes, 10,241 edges, and 2 topics. Dataset \uppercase\expandafter{\romannumeral6} consists of all original Weibo posts and related comments regarding the topics of "queue jumping by a mother and daughter," "budget travel during the holidays," and a public figure's involvement in politically sensitive issues. The data was collected from April 10, 2023, to June 10, 2023, and includes 4,005 user nodes, 14,067 edges, and 3 topics. Tabel~\ref{tab2} presents the details of the datasets. From Tabel~\ref{tab2}, we can also observe that there are not many connections between users who are interested in the same topic on online social networking platforms. This indicates that focusing solely on the connectivity between users is not sufficient; we also need to consider the possibility of information exchange between non-connected users.

\begin{table}[!t]
\caption{The Dataset\label{tab2}}
\centering
\begin{tabular}{c c c c c}
\hline
Dataset & nodes & edges & topics & \makecell{Average number of nodes \\in the initial topic} \\
\hline
Dataset \uppercase\expandafter{\romannumeral1}  & 1300 & 4951 & 1 & 15\\

Dataset \uppercase\expandafter{\romannumeral2} & 1061 & 4122 & 1 & 23\\

Dataset \uppercase\expandafter{\romannumeral3} & 1791 & 5895 & 1 & 27\\

Dataset \uppercase\expandafter{\romannumeral4} & 2300 & 8780 & 2 & 41\\

Dataset \uppercase\expandafter{\romannumeral5} & 2754 & 10241 & 2 & 45\\

Dataset \uppercase\expandafter{\romannumeral6} & 4005 & 14067 & 3 & 69\\
\hline
\end{tabular}
\end{table}

\begin{figure*}[t]
	\centering
	\subfloat[Propagation process of dataset \uppercase\expandafter{\romannumeral1}.]{\includegraphics[width=0.33\textwidth]{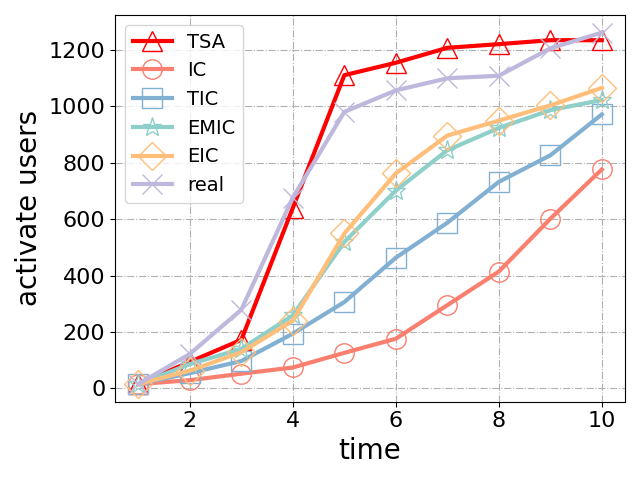}}
	\subfloat[Propagation process of dataset \uppercase\expandafter{\romannumeral2}.]{\includegraphics[width=0.33\textwidth]{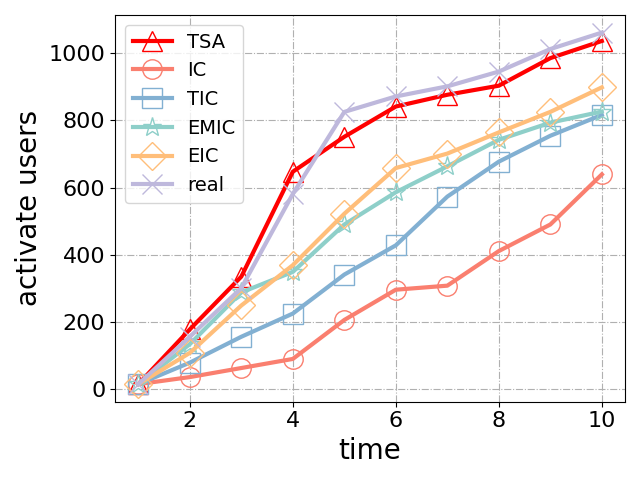}}
	\subfloat[Propagation process of dataset \uppercase\expandafter{\romannumeral3}.]{\includegraphics[width=0.33\textwidth]{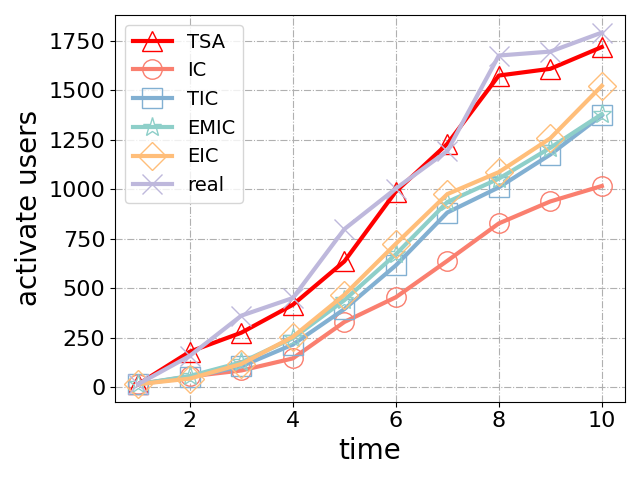}}\\
    \subfloat[Propagation process of dataset \uppercase\expandafter{\romannumeral4}.]{\includegraphics[width=0.33\textwidth]{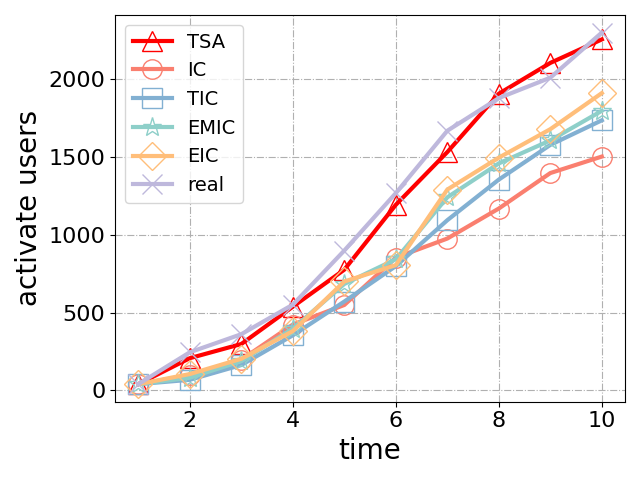}}
	\subfloat[Propagation process of dataset \uppercase\expandafter{\romannumeral5}.]{\includegraphics[width=0.33\textwidth]{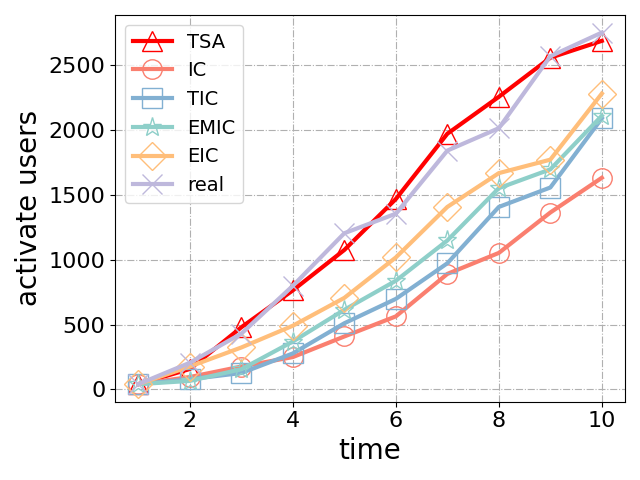}}
	\subfloat[Propagation process of dataset \uppercase\expandafter{\romannumeral6}.]{\includegraphics[width=0.33\textwidth]{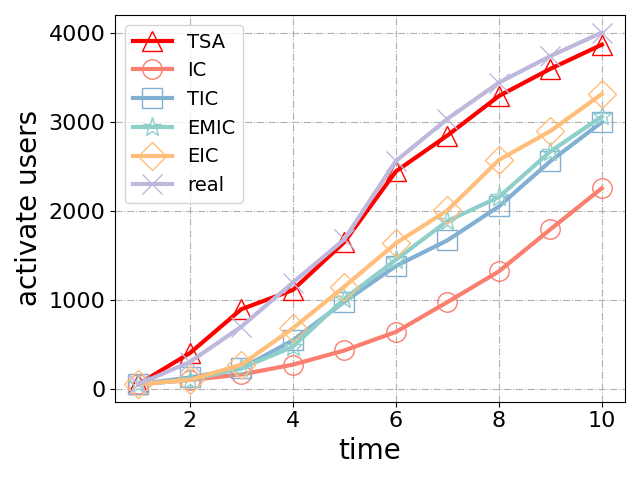}}
	\caption{The propagation process from dataset \uppercase\expandafter{\romannumeral1} to dataset \uppercase\expandafter{\romannumeral6}.}
    \label{fig2}
\end{figure*}


\subsection{Analysis of Propagation Results}
Specifically, we designate users who have posted Weibo and comments within the first half hour of each dataset as seed users for the respective topic, assuming their initial activity. The initial stance of all seed users is determined by their own Weibo posts or comments. Subsequently, we simulate the information propagation process using various propagation models.

Based on the six aforementioned real-world datasets, we conducted experiments to analyze the accuracy of the proposed model in predicting the information propagation process in real social networks. A comparison was made with traditional IC model, as well as existing models based on user sentiment or topic, namely TIC\cite{barbieri2013topic}, EMIC\cite{wang2016emotion}, and EIC\cite{dai2022opinion}. The experimental results are presented in Tabel~\ref{tab3}. It is evident that traditional propagation models, which disregard information propagation between non-adjacent nodes, struggle to accurately predict the actual information propagation process due to the limited connectivity between users in current online social networks. In contrast, the algorithm proposed in this paper takes into account the influence of all adjacent nodes, as well as selected non-adjacent nodes that may have an impact. Additionally, it considers the influence of users' different stances on the propagation process, making the spread of information more aligned with the dynamics of online social networks. The experimental results demonstrate that the proposed algorithm outperforms existing models such as IC, TIC, EMIC, and EIC in terms of accuracy. Furthermore, Fig.~\ref{fig2} illustrates the dynamic changes of the affected users when information is propagated through the TSA, IC, TIC, EMIC, and EIC models, respectively, for each of the six datasets. A comparison with the affected users in the real network further highlights the close resemblance of our model to the actual information propagation process in social networks.

\begin{table}[!t]
\caption{AUC values for IC, LT, EMIC, EIC, TIC and TSA\label{tab3}}
\centering
\begin{tabular}{c c c c c c}
\hline
Dataset & IC & TIC & EMIC & EIC & TSA \\
\hline
Dataset \uppercase\expandafter{\romannumeral1} & 61.54$\%$ & 77.01$\%$ & 78.63$\%$ & 84.48$\%$ & 97.90$\%$ \\

Dataset \uppercase\expandafter{\romannumeral2} & 61.50$\%$ & 76.92$\%$ & 77.92$\%$ & 84.64$\%$ & 98.71$\%$ \\

Dataset \uppercase\expandafter{\romannumeral3} & 59.10$\%$ & 76.68$\%$ & 77.01$\%$ & 84.97$\%$ & 97.82$\%$ \\

Dataset \uppercase\expandafter{\romannumeral4} & 65.35$\%$ & 75.49$\%$ & 78.15$\%$ & 83.12$\%$ & 97.01$\%$\\

Dataset \uppercase\expandafter{\romannumeral5} & 59.30$\%$ & 75.97$\%$ & 76.69$\%$ & 82.89$\%$ & 98.67$\%$\\

Dataset \uppercase\expandafter{\romannumeral6} & 56.28$\%$ & 74.82$\%$ & 76.43$\%$ & 82.77$\%$ & 96.62$\%$ \\
\hline
\end{tabular}
\end{table}







\begin{figure*}[t]
	\centering
	\subfloat[Stance change for dataset \uppercase\expandafter{\romannumeral1}.]{\includegraphics[width=0.33\textwidth]{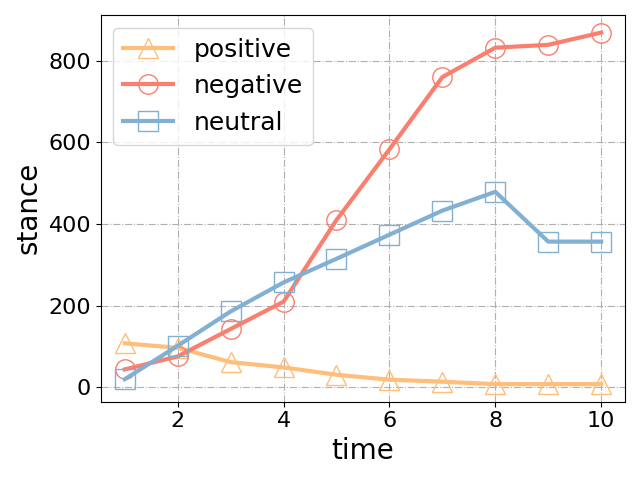}}
	\subfloat[Stance change for dataset  \uppercase\expandafter{\romannumeral2}.]{\includegraphics[width=0.33\textwidth]{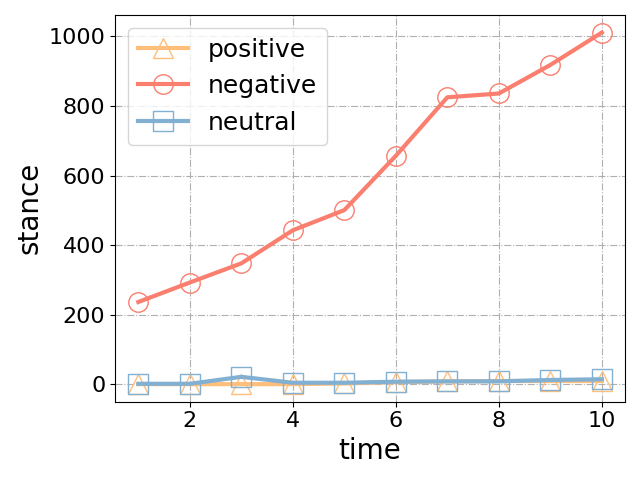}}
	\subfloat[Stance change for dataset \uppercase\expandafter{\romannumeral3}.]{\includegraphics[width=0.33\textwidth]{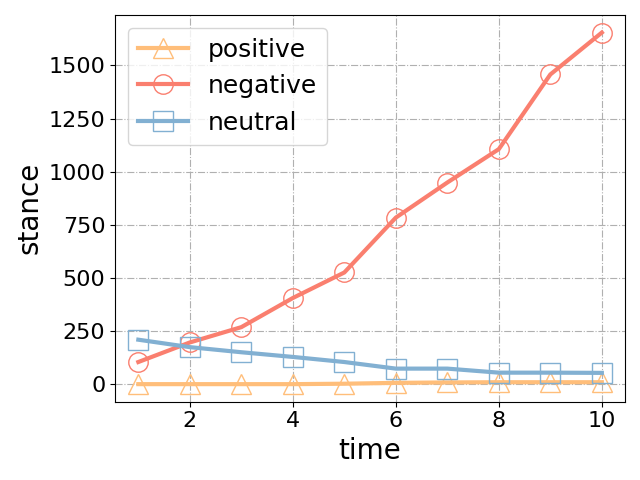}}
	\caption{The stance change process from dataset \uppercase\expandafter{\romannumeral1} to dataset \uppercase\expandafter{\romannumeral3}.}
    \label{fig3}
\end{figure*}

\subsection{Stance Change Analysis}
In this paper, three datasets containing a single topic, namely the above dataset\uppercase\expandafter{\romannumeral1}, dataset \uppercase\expandafter{\romannumeral2}, and dataset\uppercase\expandafter{\romannumeral3}, are used to analyze the stance after the dissemination, and the experimental results are shown in Tabel~\ref{tab4}. The experimental results show that our model also has high accuracy in predicting the user's stance status on the topic in information dissemination. In addition, we also conducted a dynamic analysis of the stance changes in the above three datasets during the diffusion of information dissemination, as shown in Fig.~\ref{fig3}. Obviously, in the process of information dissemination, users with different positions will continue to change, and due to the promotion of the network, users tend to have a negative attitude toward the topic of public opinion. With the continuous dissemination of information, users who were originally in a supportive or neutral position may also be influenced to change to a negative attitude.

\begin{table}[!t]
\caption{Stance prediction accuracy\label{tab4}}
\centering
\begin{tabular}{c c}
\hline
Dataset    & Stance prediction accuracy \\
\hline
Dataset \uppercase\expandafter{\romannumeral1}  & 79.40$\%$\\

Dataset \uppercase\expandafter{\romannumeral2} & 76.92$\%$ \\

Dataset \uppercase\expandafter{\romannumeral3} & 77.41$\%$\\
\hline
\end{tabular}
\end{table}

This paper conducted a stance analysis of the propagation process on three datasets that contain single topics, namely Dataset \uppercase\expandafter{\romannumeral1}, Dataset \uppercase\expandafter{\romannumeral2}, and Dataset \uppercase\expandafter{\romannumeral3}. The experimental results are presented in Tabel~\ref{tab4}. The results indicate that our model achieves high accuracy in predicting users' stances towards the topics during the information propagation process. Furthermore, dynamic analysis of stance changes in the information spreading process was conducted for the aforementioned three datasets, as depicted in Fig.~\ref{fig3}. It is evident that users with different stances undergo continuous changes during the information propagation process. Additionally, due to the influence of network dynamics, users tend to adopt a negative attitude towards public opinion topics. Moreover, as information spreads, users who initially held a supportive or neutral stance may also be influenced and shift towards a negative stance.




\section{CONCLUSION}

In the context of information propagation in online social networks, traditional information propagation models have become inadequate to adapt to the current network landscape, which involves a lot of information transmission among non-adjacent users. Therefore, this paper proposes a novel Topic and Stance-Aware model for social network information propagation and diffusion. To validate the proposed model, experiments were conducted on six real-world datasets. The experimental results demonstrate that the proposed algorithm in this paper outperforms traditional information propagation models in simulating the information diffusion process in current social networks. Furthermore, this paper analyzes the changes in users' stances towards topics during the information propagation process and dynamically illustrates the process of stance changes among users throughout the information spread. In future research, we will explore methods to select initial seed nodes that achieve the maximum propagation range while minimizing the budget requirements. Additionally, we will seek methods to promptly intercept the spread of negative information, aiming to minimize the occurrence of adverse events in the network.


%

\bibliographystyle{unsrt}
\bibliography{con}
\end{document}